\documentclass[fleqn,usenatbib]{aa}  
\usepackage{graphicx}
\usepackage{txfonts}
\usepackage{multicol}
\usepackage{color,ulem}
\usepackage{amsmath, epsfig,natbib}
\usepackage{color,ulem}
\usepackage{float}
\usepackage{newtxtext,newtxmath}
\definecolor{webgreen}{rgb}{0,.5,0}
\definecolor{webbrown}{rgb}{.6,0,0}
\usepackage[colorlinks=true,hyperfootnotes=false,%
   breaklinks=true,%
   plainpages=false, bookmarksnumbered, bookmarksopen=true,
  bookmarksopenlevel=1,%
   urlcolor=webbrown, linkcolor=blue, citecolor=blue]{hyperref}
\newcommand{\pc}{\>{\rm pc}}
\newcommand{\kpc}{\mbox{$\>{\rm kpc}$}} 

\newcommand{\Gyr}{\mbox{$\>{\rm Gyr}$}}
\newcommand{\Myr}{\mbox{$\>{\rm Myr}$}}

\newcommand\degrees{^\circ}

\begin{document} 

   \title{Looking for a needle in a haystack: Measuring the length of a stellar bar }
\titlerunning{Measuring the length of a stellar bar}
   \author{
   Soumavo Ghosh\inst{1} \thanks{E-mail: ghosh@mpia-hd.mpg.de}
      \and
   Paola Di Matteo \inst{2} \thanks{E-mail: paola.dimatteo@obspm.fr}
   }
\authorrunning{S. Ghosh and P. Di Matteo}
   \institute{Max-Planck-Institut f\"{u}r Astronomie, K\"{o}nigstuhl 17, D-69117 Heidelberg, Germany
    \and
    GEPI, Observatoire de Paris, PSL Research University, CNRS, Place Jules Janssen, 92195 Meudon, France
    }
   \date{Received XXX; accepted YYY}

  \abstract
  {
 One of the challenges related to stellar bars is to  accurately determine the length of the bar in a disc galaxy. In the literature, a wide variety of methods have been employed to measure the extent of a bar. However, a systematic study on determining the robustness and accuracy of different bar length estimators is still beyond our grasp. Here, we investigate the accuracy and the correlation (if any) between different bar length measurement methods while using an $N$-body model of a barred galaxy, where the bar evolves self-consistently in the presence of a live dark matter halo. We investigate the temporal evolution of the bar length, using different estimators (involving isophotal analysis of de-projected surface brightness distribution and Fourier decomposition of surface density), and we study their robustness and accuracy. We made further attempts to determine correlations among any two of these bar length estimators used here. In the presence of spirals, the bar length estimators that only consider the amplitudes of different Fourier moments (and do not take into account the phase-angle of $m=2$ Fourier moment) systematically overestimate the length of the bar. The strength of dark-gaps (produced by bars)  is strongly correlated with the bar length in early rapid growth phase and is only weakly anti-correlated during subsequent quiescent phase of bar evolution. However, the location of dark-gaps is only weakly correlated to the bar length, hence, this information cannot be used as a robust proxy for determining the bar length. In addition, the bar length estimators, obtained using isophotal analysis of de-projected surface brightness distribution, systematically overestimate the bar length. The implications of bar length over(under)estimation in the context of determining fast and slow bars are further discussed in this work.

   }

    \keywords{galaxies: kinematics and dynamics - galaxies: structure - galaxies: spiral - galaxies: evolution - methods: numerical }

   \maketitle
%

\section{Introduction}
\label{sec:Intro}

Stellar bars are one of the most common non-axisymmetric structures in disc galaxies in the local Universe \citep[e.g. see][]{Eskridgeetal2000,NairandAbraham2010,Mastersetal2011,Kruketal2017}. The Milky Way also hosts a stellar bar in the central region \citep[e.g.][]{LisztandBurton1980,Binneyetal1991,Weinberg1992,Binneyetal1997,BlitzandSpergel1991,Hammersleyetal2000,WegandGerhard2013}. Bars are present in high redshift ($z \sim 1$) disc galaxies as well with the bar fraction decreasing with redshift \citep[e.g. see][but also see \citet{Elmetal2004,Jogeeetal2004}]{Shethetal2008,Melvinetal2014,Simmonsetal2014}. Furthermore, recent studies by \citet{Guoetal2022}, \citet{LeConteetal2023}, and \citep{Costantinetal2023} showed the presence of conspicuous stellar bars even at a higher redshift ($z \sim 1.1-3$). At those redshifts, the discs are known to be thick, kinematically hot (and turbulent), and more gas-rich; therefore, a tidally induced origin has been proposed to explain the presence of bars in such high-redshift galaxies. However, recent work by \citet{Ghoshetal2023} demonstrated that bars can form purely from an internal gravitational instability, even in presence of a massive thick disc (analogous to those high redshift galaxies). 
\par
Dedicated theoretical efforts have been made in past decades to understand the formation and evolution of stellar bar as well as to understand secular evolution driven by bars \citep[e.g.][]{CombesandSanders1981,SellwoodandWilkinson1993,DebattistaSellwood2000,Athanassoula2003}.  Furthermore, understanding the variation of bar properties with host galaxy's properties (e.g. stellar mass, Hubble type) has proven to be instrumental for gaining insight into the formation and evolutionary scenario of bars \citep[e.g. see][]{Kormdendy1979}. There are three properties of bars which are of key interest in galactic dynamics (and can be measured from observations), namely, the strength of the bar, the pattern speed of the bar, and the length of the bar. 
\par
For both the observed and the simulated barred galaxies, the strength of a bar is measured by using different techniques. To mention a few, the bar strength is quantified from the amplitude of the $m=2$ Fourier moment \citep[e.g.][]{Aguerrietal1999,Aguerrietal2000, DebattistaSellwood2000,Athanassoula2003,Ghoshetal2021,Ghoshetal2023}, using a two-dimensional fast Fourier transform (FFT) method \citep{Garciaetal2017}, from the ratio of the peak surface brightness in the bar to the minimum surface brightness in the inter-bar region \citep[e.g.][]{ElmeandElme1985,Reganetal1997} by measuring the bar torque, $Q_{b}$ \citep[e.g.][]{CombesandSanders1981,ButaandBlock2001,LaurikainenandSalo2002} and, then, from the ellipticity of the best-fitted elliptical isophotes \citep[e.g.][]{Aguerrietal1999,Shlosmanetal2000,MarinovaandJogee2007}. On the other hand, the bar pattern speed ($\Omega_{\rm bar}$) is measured either via indirect methods, for example, by locating the sign-reversal of streaming motions across the corotation radius, by associating the rings with the Outer Lindblad resonance of the bar \citep[e.g.][]{Buta1986,Jeongetal2007} or by directly measuring from the disc kinematics \citep[e.g.][]{MerrifieldKuijken1995,Gressenetal1999,Gressenetal2003,vpdetal2004,Aguerrietal2015,Cuomoetal2019,Guoetal2019} using the Tremaine-Weinberg method \citep{TremaineWeinberg1984}. However, to denote whether a bar is rotating faster or slower, the parameter $\mathcal{R}$, defined as the ratio of the bar corotation radius ($R_{\rm CR}$) and the bar length ($R_{\rm bar}$), is widely used in the literature, where $ 1 < \mathcal{R} < 1.4$ denotes a fast bar, $\mathcal{R} >1.4$ denotes a slow bar, and $\mathcal{R} < 1 $ denotes a ultra-fast bar \citep[for details, see][]{DebattistaSellwood2000}.
\par
One of the longstanding problems related to bars is determining the extent of the bar or where the bar ends morphologically in a real galaxy. Since a stellar bar is a continuous structure, often embedded in a conspicuous stellar bulge, the bar length often crucially depends on the method of bar length estimation itself. In the literature, a wide variety of methods or techniques has been employed to determine the length or extent of a bar. To illustrate, the bar length is quantified from the structural decomposition of radial surface brightness profiles \citep[e.g.][]{deJong1996,Prietoetal1997,Aguerrietal2003,Gadotti2011,Kruketal2017} by visual inspection \citep{Martin1995}, from the peak of the isophotal ellipticity \citep[e.g. see][]{Wozniaketal1995,Erwin2005,Aguerrietal2009}, from variations of the Fourier modes phase angle of the galaxy light distribution \citep{Quillenetal1994,Aguerrietal2003} by measuring the intensity ratio along the bar and inter-bar region \citep[e.g.][]{Ohtaetal1990,Aguerrietal1998,Aguerrietal2000}, {from the azimuthally averaged radial profile of the transverse-to-radial force ratio \citep{Leeetal2020}}, and from the constancy of the isophotal position angle \citep[e.g.][]{Sheth2003,Erwin2005} by measuring the radial location where the $m=2$ Fourier coefficient drops to a certain fraction of its peak value \citep[e.g.][]{Fragkoudietal2021,Ghoshetal2023} or by manually placing a bar-shaped masking on the SDSS images \citep{Mastersetal2021}. In addition, a recent work by \citet{Petersenetal2023} demonstrated a technique of measuring the bar length by calculating the maximum extent of the most stable $x_1$-orbits in the underlying potential. Another recent work by \citet{Luceyetal2023} showed that an estimate of the bar length can be obtained based on the maximal extent of trapped bar orbits, computed in an assumed galactic potential. While these methods rely on robust dynamical arguments, the resulting bar length measurement critically depends on the assumed underlying potential of the galaxy. 
\par
The importance of measuring the bar length extends beyond the usual notion of small and large bars in different disc galaxies. \citet{Kormdendy1979} showed, for the first time, that bar length is correlated with the luminosity of host galaxies. In addition, bar lengths are aptly correlated with the Hubble type of the host galaxies, such that early-type disc galaxies tend to host larger bars as compared to their late-type counterparts \citep[e.g. see][]{ElmeandElme1985,Reganetal1997,Menéndez-Delmestre_2007,Aguerrietal2009}. Further correlations of bar length with host galaxy mass, host galaxy colour, disc scale length, and bulge size have been studied in the literature \citep[e.g. see][]{Aguerrietal2005,MarinovaandJogee2007,Gadotti2011}. Taken altogether, it outlines the importance of studying the bar length to better understand the secular evolution of disc galaxies. Furthermore, past studies have shown that the tidally induced bars are generally larger than those formed via internal gravitational instability \citep[e.g. see][]{ValenzuelaandKlypin2003,Holley-Bockelmann2005}; thereby providing important clues about past merger histories of the host galaxy. The bar length measurement holds a crucial role in determining a fast and slow bar as well. A systematic overestimation in bar length measurement can erroneously map an otherwise slow bar in the `fast bar' regime (by lowering the value of $\mathcal{R}$) or even an otherwise fast bar in an `ultra-fast bar' regime \citep[for details, see e.g.][]{Hilmietal2020,Cuomoetal2021}.
\par
So far, a systematic comparative study of different bar length estimators is largely missing in past literatures \citep[but see discussions in][]{Fragkoudietal2021,Petersenetal2023}. Furthermore, it still remains unclear whether any (positive) correlation exists between different bar length estimators. {\citet{AthanassoulaandMisiriotis2002} studied different estimators for the bar length measurement from three $N$-body models of barred galaxy. In addition,} past study by \citet{Michel-Dansac2006} investigated the accuracy and reliability of six methods (intrinsic and/or photometric) used to determine the length of stellar bars in a simulated galaxy as well as in observed galaxies. However, their simulated galaxy did not reveal a live dark matter halo, thereby casting doubt on their inferred temporal evolution of $\mathcal{R,}$ as well as the shape and the strength of the bar. Hence, a similar investigation needs to be carried out where the bar evolves self-consistently in a live dark matter halo.  We aim to address these issues here.
\par
In this paper, we systematically compute the bar length, obtained by using different estimators  commonly used in past literature, as a function of time for an $N$-body model of a barred galaxy. In addition, we quantify the correlations between different bar length estimators, obtained from the photometry and the intrinsic stellar distribution. Furthermore, the effect of spirals on introducing systematic biases in bar length measurements and the correlation between the extent of the dark lanes and bar length are studied in detail.
\par
The rest of the paper is organised as follows.
Section~\ref{sec:details_of_the_model} provides a brief description of the  initial equilibrium simulation set-up and the bar evolution in our fiducial model. Section~\ref{sec:bar_measurement} contains the details of temporal evolution of bar lengths obtained from different estimators and the correlation study between them, while Sect.~\ref{sec:dark_lanes} provides the details of dark-lane properties and their variation with bar length. 
Section~\ref{sec:bar_spiral_conundrum}  contains the results of spirals' effect on bar length measurements, while Sect.~\ref{sec:comparison_photometry} provides the details of correlations between different intrinsic and photometric bar length estimators. Section~\ref{sec:discussion} contains the discussion and  Sect.~\ref{sec:conclusion} summarises the main findings of this work.

\section{$N$-body model of a barred galaxy}
\label{sec:details_of_the_model}

To motivate the study, we consider an $N$-body model of a collisionless stellar disc that subsequently forms a prominent bar. We use this fiducial bar model to test the reliability and robustness of various bar length estimators, commonly used in the literature.
\subsection{Simulation set-up and equilibrium configuration}
\label{sec:sim_setup}
 Here, for the sake of completeness, we briefly mention the equilibrium set-up for our fiducial bar model. For details of the initial equilibrium model and how they are generated, we refer to \citet{Ghoshetal2023} (see, in particular, the rthick0.0 model there).
\par
The equilibrium model is made up of an axisymmetric stellar disc embedded in a live dark matter halo. The stellar disc is modelled with a Miyamoto-Nagai profile \citep{MiyamatoandNagai1975}, featuring a characteristic disc scale length of $R_{\rm d} = 4.7 \kpc$, a scale height of $h_{\rm z} = 0.3 \kpc$, and a total disc mass of $M_{\rm disc} = 1\times 10^{11} M_{\odot}$. The dark matter halo is modelled with a Plummer sphere \citep{Plummer1911}, featuring a characteristic scale length of $R_{\rm H} = 10 \kpc$ and  a total dark matter halo mass of $M_{\rm dm} = 1.6 \times 10^{11} M_{\odot}$. A total of $1 \times 10^6$ particles are used to model the stellar disc, while a total of $5 \times 10^5$ particles are used to model the dark matter halo.
\par
The initial condition of the stellar disc was obtained using an iterative method algorithm \citep[see][]{Rodionovetal2009}. For further details, we refer to \citet{Ghoshetal2023}. The simulation was run, using a TreeSPH code by \citet{SemelinandCombes2002}, for a total time of $9 \Gyr$. A hierarchical tree method \citep{BarnesandHut1986} with an opening angle of $\theta = 0.7$ was used for calculating the gravitational force, which includes terms up to the quadrupole order in the multipole expansion. Furthermore, a Plummer potential is used for softening the gravitational forces with a softening length of $\epsilon = 150 \pc$.

\begin{figure*}
\includegraphics[width=1.02\linewidth]{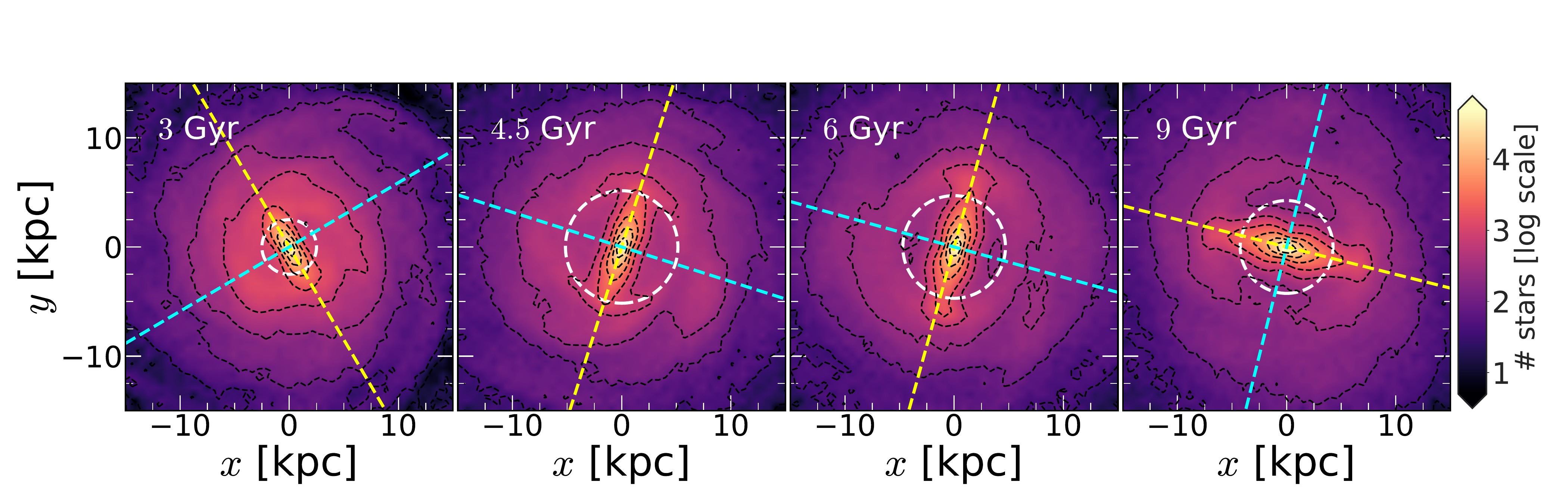}
\caption{Face-on stellar density distribution at four different times, denoting different phases of the bar evolution (for details, see the text). The black dashed lines denote the contours of constant density. The yellow dashed lines denote the bar major-axis whereas the cyan dashed lines denote the bar minor-axis, respectively. The white dashed circle denotes the bar length ($R_{\rm bar,1}$), measured from the constancy of the phase-angle of the $m=2$ Fourier moment. For details, we refer to the text.}
\label{fig:density_maps}
\end{figure*}

\begin{figure*}
\includegraphics[width=1.0\linewidth]{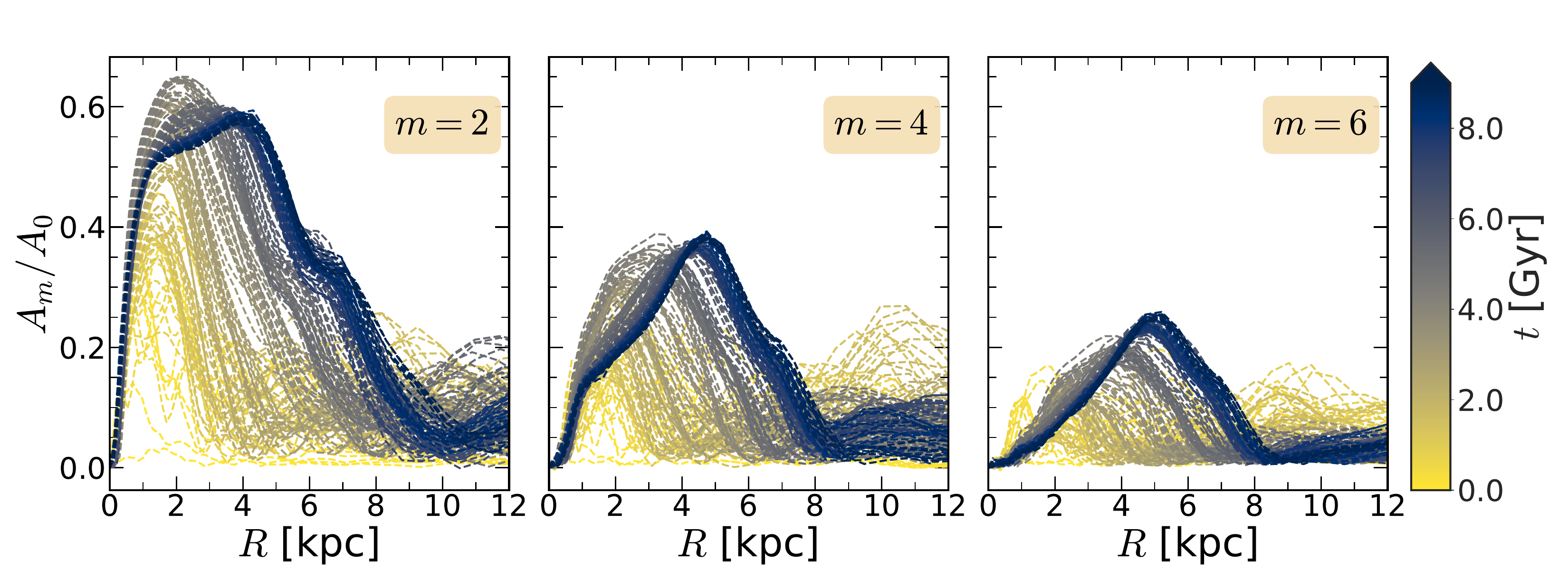}
\caption{Radial variation of the $m=2, 4, 6$ Fourier coefficients (normalised by the $m=0$ component), computed using Eq.~\ref{eq:fourier_calc}, as a function of time (shown in the colour bar) for the fiducial bar model. }
\label{fig:fourier_radial_profiles}
\end{figure*}

\subsection{Temporal evolution of the bar model}
\label{sec:barred_model}

Figure~\ref{fig:density_maps} shows the stellar density distribution in the face-on projection at four different times, denoting the early rapid bar growth, a fully grown bar, a bar undergoing the vertical buckling instability, and the final stage of the bar by the end of the simulation run. To quantify the bar strength, we calculate the radial variations of different azimuthal Fourier components of the mass distribution in the disc using \citep[for details, see][]{Ghoshetal2023}:
\begin{equation}
A_m/A_0 (R)= \left| \frac{\sum_i m_i e^{im\phi_i}}{\sum_i m_i}\right|\,,
\label{eq:fourier_calc}
\end{equation}
\noindent where $A_m$ denotes the coefficient of the $m^{th}$ Fourier moment of the density distribution, $m_i$ is the mass of the $i^{th}$ particle, and $\phi_i$ is the corresponding cylindrical angle. A strong bar ($m=2$ Fourier moment) often induces an $m=4$ Fourier moment (albeit weaker than the $m=2$ mode), whereas a face-on boxy/peanut (hereafter b/p) structure can be represented by an $m=6$ Fourier moment \citep{CiamburGraham2016,Sahaetal2018}. Fig.~\ref{fig:fourier_radial_profiles} shows the corresponding radial profiles of the $m =2, 4, 6$ Fourier coefficients as a function of time. A prominent peak in the $A_2/A_0$ radial profiles, accompanied by a similar peak (but with a lower peak value) in the $A_4/A_0$ radial profiles clearly confirms the presence of a strong, central stellar bar in our fiducial model. The bar forms in our fiducial model around $t \sim 1.1 \Gyr$. For further details of the temporal evolution of the bar strength, the reader is referred to \citet{Ghoshetal2023}. Around $t = 4.55 \Gyr$, the bar undergoes a vertical buckling instability, thereby forming a prominent b/p structure at later times \citep{Ghoshetal2023b}. In addition, we measured the  bar pattern speed ($\Omega_{\rm bar}$) for our fiducial bar model. The details are shown in Appendix~\ref{appen:vcirc_calc}. The bar slows down significantly by the end of the simulation run ($t = 9 \Gyr$).

\section{Measurements of length of the bar}
\label{sec:bar_measurement}

Here, we systematically employ different estimators for measuring the length of a bar, as a function of time, in our fiducial model. For this work, we selected five such bar length estimators obtained from the intrinsic particle distribution, and then studied the correlation between them. As the bar evolves through different phases (e.g. rapid growth, buckling instability) with time; therefore, studying such correlations (using snapshots at different times) would mimic a sample of observed barred galaxies with bars in their different evolutionary phases. In addition, our choice of these five bar length estimators is motivated by the fact that these measurements can  be aptly carried out for a large sample of observed barred galaxies, thereby leaving the future scope of testing the robustness of the correlations reported here. Below, we list the bar length measurement methods, commonly used in past literatures, employed in this work.

\begin{itemize}

\item{$R_{\rm bar,1}$ : the radial extent within which the phase-angle of the $m=2$ Fourier moment ($\phi_2$) remains constant within $\sim 3-5 \degrees$;}

\item{$R_{\rm bar,2}$ : the radial extent where the amplitude of the $m=2$ Fourier moment ($A_2/A_0$) attains its maximum value;}

\item{$R_{\rm bar,3}$ : the radial extent where the amplitude of the $m=2$ Fourier moment ($A_2/A_0$)  drops to 70 percent of its peak value;}

\item{$R_{\rm bar,4}$ : the radial extent where the ratio of the bar intensity ($I_b$) and the inter-bar intensity ($I_{ib}$) exceeds 2, that is, $I_b/I_{ib} >2$;}

\item{$R_{\rm bar,5}$ : the radial extent where the ratio of the bar intensity ($I_b$) and the inter-bar intensity ($I_{ib}$) obeys the following inequality $\left( I_b/ I_{ib}\right) >  \left( I_b/ I_{ib}\right)_{\rm min} +1/2 [ \left( I_b/ I_{ib}\right)_{\rm max} -  \left( I_b/ I_{ib}\right)_{\rm min}]$.}
\end{itemize}

\begin{figure*}
\centering
\resizebox{1.02\linewidth}{!}{\includegraphics{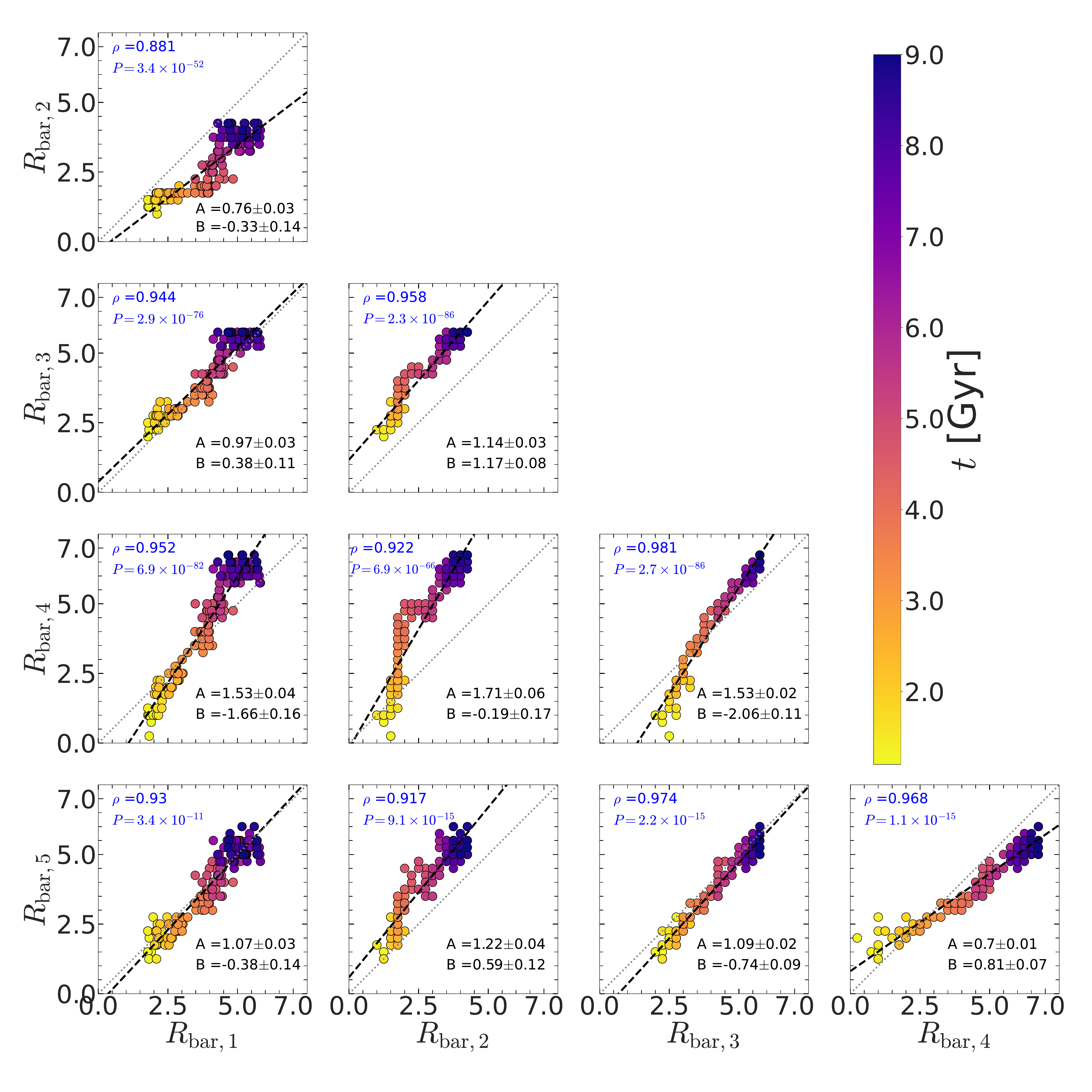}}
\caption{Comparison of different methods of measuring the length of the bar, as a function of time (see the colour bar) for the fiducial bar model. For details of the bar measurement methods, see Sect.~\ref{sec:bar_measurement}. In each case, a straight line of the form $Y = AX+B$ is fitted (black dashed line) and the corresponding best-fit parameters are quoted in each sub-panel. In addition, in each case, the Pearson correlation is computed, and they are quoted in each sub-panel (see top left). The grey dotted line in each sub-panel denotes the one-to-one correspondence:}
\label{fig:bar_length_measure}
\end{figure*}

Following \citet{Aguerrietal2000}, we define the bar and the interbar intensity via the $m=2, 4, 6$ Fourier amplitudes using:
\begin{eqnarray}
I_b (R) = I_0 (R)+ I_2 (R)+I_4(R)+ I_6(R),\\\nonumber
I_{ib}  (R)= I_0(R)- I_2(R)+I_4(R)- I_6(R)\,,\\\nonumber
\end{eqnarray}
\noindent where $I_m$ is the amplitude of the $m$th Fourier moment. Figure~\ref{fig:bar_length_measure} shows the corresponding temporal evolution of the bar length, measured using the five estimators described above. We also studied if there is any correlation between any two of these bar length measurement estimators. These are also shown in Fig.~\ref{fig:bar_length_measure}. For this work, we used the bar length, $R_{\rm bar,1}$ as the benchmark, and then we compared other bar length estimators to see whether they introduce any systematic overestimation or underestimation in the bar length measurement. However, we mention that, for each pair constructed from the five estimators, a best-fit linear relation is computed. This, in turn, will help one to use any other bar length estimator as a 'benchmark'. Below, we list the main findings of the bar length measurements and their associated correlations. We computed the Pearson correlation coefficient throughout this paper to quantify the correlation between any two estimators of the bar length.

\begin{itemize}

\item{The bar length, $R_{\rm bar,2}$, obtained from the peak location of the $m=2$ Fourier coefficient, is well-correlated with $R_{\rm bar,1}$ with the Pearson correlation coefficient, $\rho = 0.88$. However, the values of $R_{\rm bar,2}$ systematically underestimate the bar length determined by $R_{\rm bar,1}$, as is evident from the best-fit slope of the fitted straight line.}

\item{The bar length, $R_{\rm bar,3}$ is strongly correlated with $R_{\rm bar,1}$ with the Pearson correlation coefficient, $\rho = 0.94$. In addition, the values of $R_{\rm bar,3}$ display an (almost) one-to-one correspondence with the bar extent determined by $R_{\rm bar,1}$, as revealed by the best-fit slope of the fitted straight line.}

\item{The bar length, $R_{\rm bar,4}$ is strongly correlated with $R_{\rm bar,1}$ as well with the Pearson correlation coefficient, $\rho = 0.95$. However, the values of $R_{\rm bar,4}$ systematically overestimate the bar extent determined by $R_{\rm bar,1}$, as is evident from the best-fit slope of the fitted straight line.}

\item{{The bar length, $R_{\rm bar,5}$ is also strongly correlated with $R_{\rm bar,1}$ with the Pearson correlation coefficient, $\rho = 0.93$. Furthermore, the values of $R_{\rm bar,3}$ display an (almost) one-to-one correspondence with the bar extent determined by $R_{\rm bar,1}$, as revealed by the best-fit slope of the fitted straight line\footnote{We mention that in the early bar growth phase, for about four to five snapshots, the values of $R_{\rm bar,5}$ were very large (and unphysical). If we include these points, the corresponding correlation values will decrease and also will result in a worse best-fit values. Therefore, in Fig.~\ref{fig:bar_length_measure}, we excluded these points. Since these four to five points do not correspond to any specific time interval in the bar evolution and that they are redistributed over the $5 \Gyr$ of the simulation, therefore, their removal does not affect the physical interpretation of our findings.}.}}

\end{itemize}

{The finding of consistent bar length measurement as quantified by $R_{\rm bar,5}$ is in agreement with the past findings of \citet{Michel-Dansac2006} and \citet{AthanassoulaandMisiriotis2002} who found the least noisy bar length measurement by this method. However, we mention that at later times ($t > 5 \Gyr$), the values of $R_{\rm bar,5}$ systematically lie above(under) the best-fit straight line denoting the correlation between $R_{\rm bar,1}$ and $R_{\rm bar,5}$ (see bottom left panel in Fig.~\ref{fig:bar_length_measure}). In Appendix.~\ref{appen:dissection_rbar5}, we study this in further detail.}
\par
{We further mention that, at initial rapid bar growth phase, the values of $R_{\rm bar, 2}$ do not increase appreciably (as compared to other bar length estimators), thereby producing a non-linear, step-like behaviour (see second column of Fig.~\ref{fig:bar_length_measure}). We checked that at those epochs, the peak value of the $m=2$ Fourier coefficient increases substantially (denoting the rapid growth in bar strength), without changing the radial location where the peak occurs (i.e. value of $R_{\rm bar,2}$). This trend can also be judged from the radial profiles of the $m=2$ Fourier coefficient shown in Fig.~\ref{fig:fourier_radial_profiles}.
\par
Lastly, our assumption of the phase-angle of the $m=2$ Fourier moment ($\phi_2$) being constant within $3-5 \degrees$ in the bar region, is somewhat a stricter condition imposed as compared to past studies where a constancy of $\phi_2$ within $\sim 10 \degrees$ were assumed. The motivations behind choosing a stricter condition are the following. This choice helps to avoid any overestimation from the spirals at the early epochs where the model harbours both a bar and prominent  spirals. In addition, \citet{Vynatheyaetal2021} demonstrated that when a strong bar subsequently forms a face-on peanut (similar to the model considered here), the values of $\phi_2$ deviate by $\sim 8-10 \degrees$ in the outer region encompassing the `ansae' or the `handle' of the bar (see Fig. 3 there).  It is still not clear whether the outer `ansae' of the bar is supported by the same family of $x_1$-orbits which constitute the inner bar.  Therefore, our choice of $\phi_2$ within $3-5 \degrees$ in the bar region leaves the ansae part of the bar later times, as also can be judged visually from Fig.~\ref{fig:density_maps} (compare the white circles and the contours of the constant density).

\section{Quantifying the light deficit around the bar}
\label{sec:dark_lanes}

\begin{figure*}
\centering
\resizebox{0.9\linewidth}{!}{\includegraphics{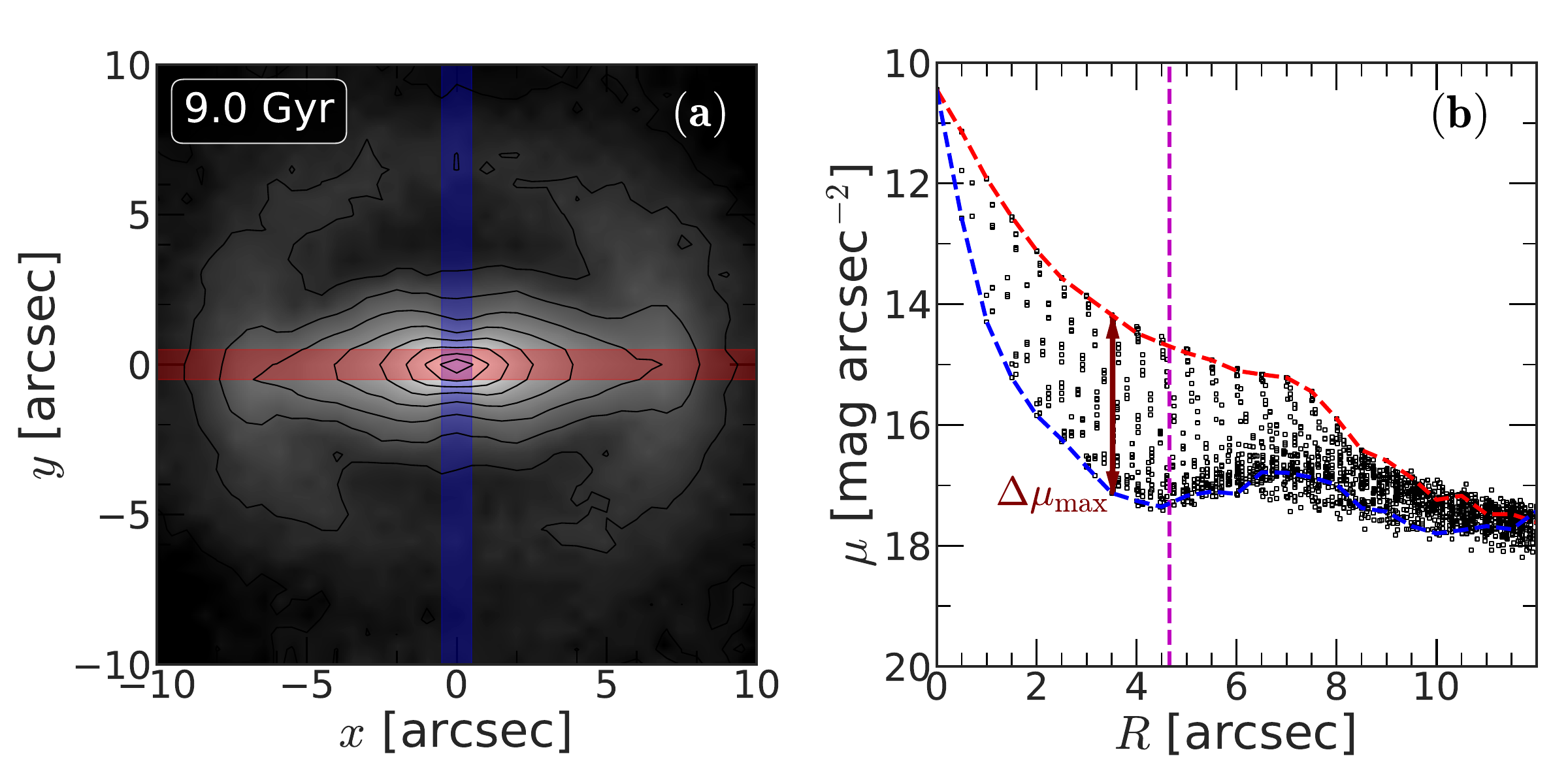}}
\medskip
\resizebox{0.9\linewidth}{!}{\includegraphics{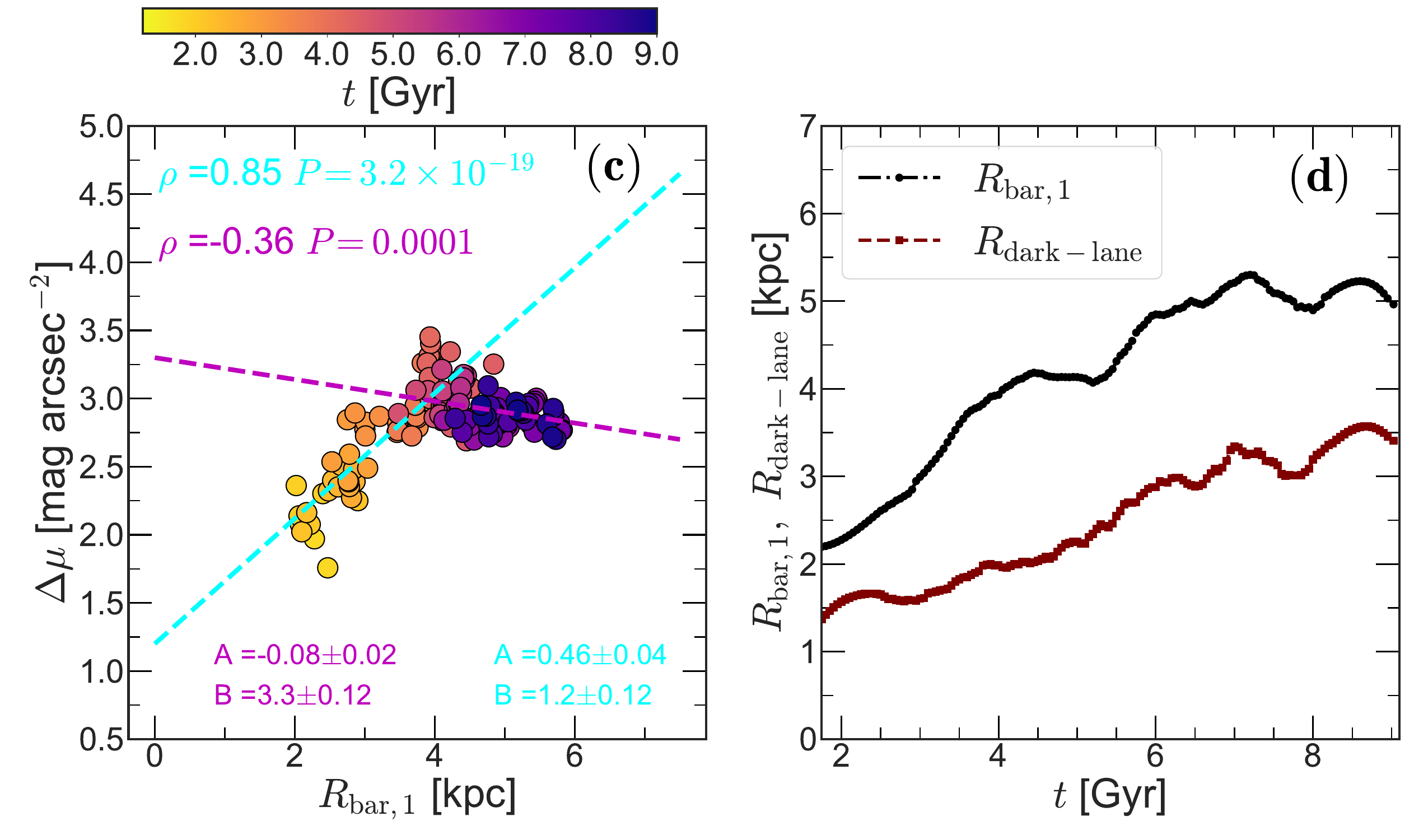}}
\caption{Dark-lanes as a diagnostic for bar length measurement. \textit{Panel (a):} Face-on surface brightness distribution at $t = 9 \Gyr$. The solid black lines denote the contours of constant surface brightness. Here, a conversion of $1$ arcsec = $1 \kpc$ and a magnitude zero-point ($m_0$) of $22.5$ mag arcsec$^{-2}$ are used to create the surface brightness map from the intrinsic particle distribution. The red and blue lines denote the bar's major  and minor axis, respectively.  \textit{Panel (b):} Corresponding light profile along the the bar major and minor axis (red and blue dashed lines, respectively). The radial location where the light deficit around the bar reaches its maximum is indicated by the maroon arrow. The vertical magenta line denote the bar length, $R_{\rm bar,1}$. A solid square represents a single pixel of the face-on surface brightness map. \textit{Panel (c):} Distribution of the maximum  light deficit around the bar with the bar length, $R_{\rm bar,1}$ (obtained from the constancy of the $m=2$ phase-angle). A straight line of the form $Y = AX+B$ is fitted separately (cyan and magenta dashed lines) to points, before and after $t = 5 \Gyr$, and the corresponding best-fit values are quoted. For details, see the text. The Pearson correlations are computed, and shown in the top left corner. \textit{Panel (d):} Temporal evolution of $R_{\rm bar,1}$ (bar length) and $R_{\rm dark-lane}$ for the fiducial bar model. For further details, see the text.}
\label{fig:darklanes}
\end{figure*}

Past theoretical works have shown that as a bar grows in size and mass, it continuously traps stars that are on nearly-circular orbits onto the $x_1$ orbits that serves as a backbone for the bar structure \citep[e.g. see][]{ContopoulosandGrosbol1989,Athanassoula2003,BiineyTremaine2008}. This makes an otherwise azimuthally-smooth light profile into a rather radially bright light profile; thereby causing a light deficit along the bar minor axis within the extent of the bar. This deficit in the light profile along the bar minor axis is seen in observed barred galaxies as well \citep[e.g. see][]{Gadottietal2003,Laurikainenetal2005,Gadotti2008,Kimetal2015}. A recent study by \citet{Krishnaraoetal2022} showed that these `dark-gaps' are associated with the location of the 4:1 ultraharmonic resonance of the bar instead of the bar corotation, {further confirmed in a recent study by \citet{Aguerrietal2023}}. \citet{Kimetal2016} studied the variation of this light deficit with different bar properties (e.g. bar length, Bar/T) using a sample of barred galaxies from the \textit{Spitzer} Survey of Stellar Structure in Galaxies (S$^4$G). However, a systematic study of variation of the light deficit with the bar length is still missing \citep[{but see}][{for a recent exploration of dark-gaps with bar length}]{Aguerrietal2023}. Furthermore, the question of whether the location of maximum light deficit is correlated with the bar length is still not well understood. We address these issues below.
\par
First, we checked that as a bar grows in strength and size in our fiducial model, it creates  a prominent light deficit along the bar minor axis. To study this in further detail, we then computed the face-on surface brightness distribution from the intrinsic particle distribution while using a conversion of 1 arcsec = $1 \kpc$ and a magnitude zero-point ($m_0$) of $22.5$ mag arcsec$^{-2}$.  In each case, the bar is rotated in such a way that it always remains aligned with the $x$-axis. One such example, computed at $t = 9 \Gyr$ is shown in Fig.~\ref{fig:darklanes} (see panel (a)). Next, we extracted the average surface brightness profiles along the bar major- and minor-axis, while putting a slit of width, $\Delta = 0.5 \kpc$ in each direction (see panel (b) in Fig.~\ref{fig:darklanes}). To study whether there is any correlation between the dark lane's properties and the bar's properties, we calculated the maximum light deficit, $\Delta \mu_{\rm max}$, and the radial location where $\Delta \mu_{\rm max}$ occurs. The resulting variation of maximum light deficit, $\Delta \mu_{\rm max}$ with bar length, $R_{\rm bar,1}$ (obtained via a constancy of $\phi_2$ phase-angle) as a function of time for our fiducial model is shown in Fig.~\ref{fig:darklanes} (see panel (c)). When all the points are taken together, the best-fit straight line as well as the values of the Pearson correlation coefficients ($\rho = 0.58$) indicate that  these two quantities are only weakly correlated. From the S$^4$G sample, \citet{Kimetal2016} found a slightly stronger correlation between these two quantities (Pearson correlation coefficient, $\rho \sim 0.68-0.75$). However, a visual inspection of panel (c) of Fig.~\ref{fig:darklanes} seems to indicate that these two quantities are strongly related in the earlier times ($\sim 5 \Gyr$ or before) as opposed to the later times. To investigate this further, we divided the points into two parts, namely, before and after $t = 5 \Gyr$, and fitted straight-lines to these two datasets separately. This is shown in Fig.~\ref{fig:darklanes} (see panel (c)). As seen clearly from the best-fit straight line as well as the values of the Pearson correlation coefficients ($\rho = 0.85$), indeed, the maximum light deficit, $\Delta \mu_{\rm max}$, is strongly correlated with bar length, $R_{\rm bar,1}$, at earlier times ($t \leq 5 \Gyr$) as compared to the later stages ($t > 5 \Gyr$), where these two quantities are weakly anti-correlated ($\rho = -0.36$). The physical reason behind these two trends at different epochs is the following. Within the first $5 \Gyr$ time, the bar evolves quite strongly, going through the early rapid growth phase and undergoing the vertical buckling instability. However, after $t = 5 \Gyr$ or so, the bar in our fiducial model does not grow appreciably in strength, and the bar length increases only moderately \citep[for further details, see][]{Ghoshetal2023}. As the maximum light deficit, $\Delta \mu_{\rm max}$ strongly correlates with the bar strength, therefore, the different evolutionary pathways (before and after $5 \Gyr$) result in two strikingly different trends for the correlations between the maximum light deficit and the bar length.
 
  Lastly, we define $R_{\rm dark-lane}$ as radial location where $\Delta \mu_{\rm max}$ occurs at time, $t$. The resulting temporal variation of the $R_{\rm dark-lane}$ and the bar length ($R_{\rm bar,1}$) for our fiducial model is shown in Fig.~\ref{fig:darklanes} (see panel (d)). The temporal evolution of $R_{\rm dark-lane}$ broadly follows the temporal evolution of bar length. This is consistent with the dark-lane formation scenario by trapping more stars in radially-elongated $x_1$-orbits. However, we did not find any (strong) correlation between $R_{\rm dark-lane}$ and the length of the bar, in our fiducial model. This implies that the radial location of the maximum light deficit can not be used as a robust proxy for the bar length estimations. We further checked that the $R_{\rm dark-lane}$ values do not correspond to the bar corotation, in agreement with the findings of \citet{Krishnaraoetal2022}. For the sake of brevity, these are not shown here.

\section{Influence of spiral arms in measuring the bar length}
\label{sec:bar_spiral_conundrum}

\begin{figure*}
\centering
\resizebox{\linewidth}{!}{\includegraphics{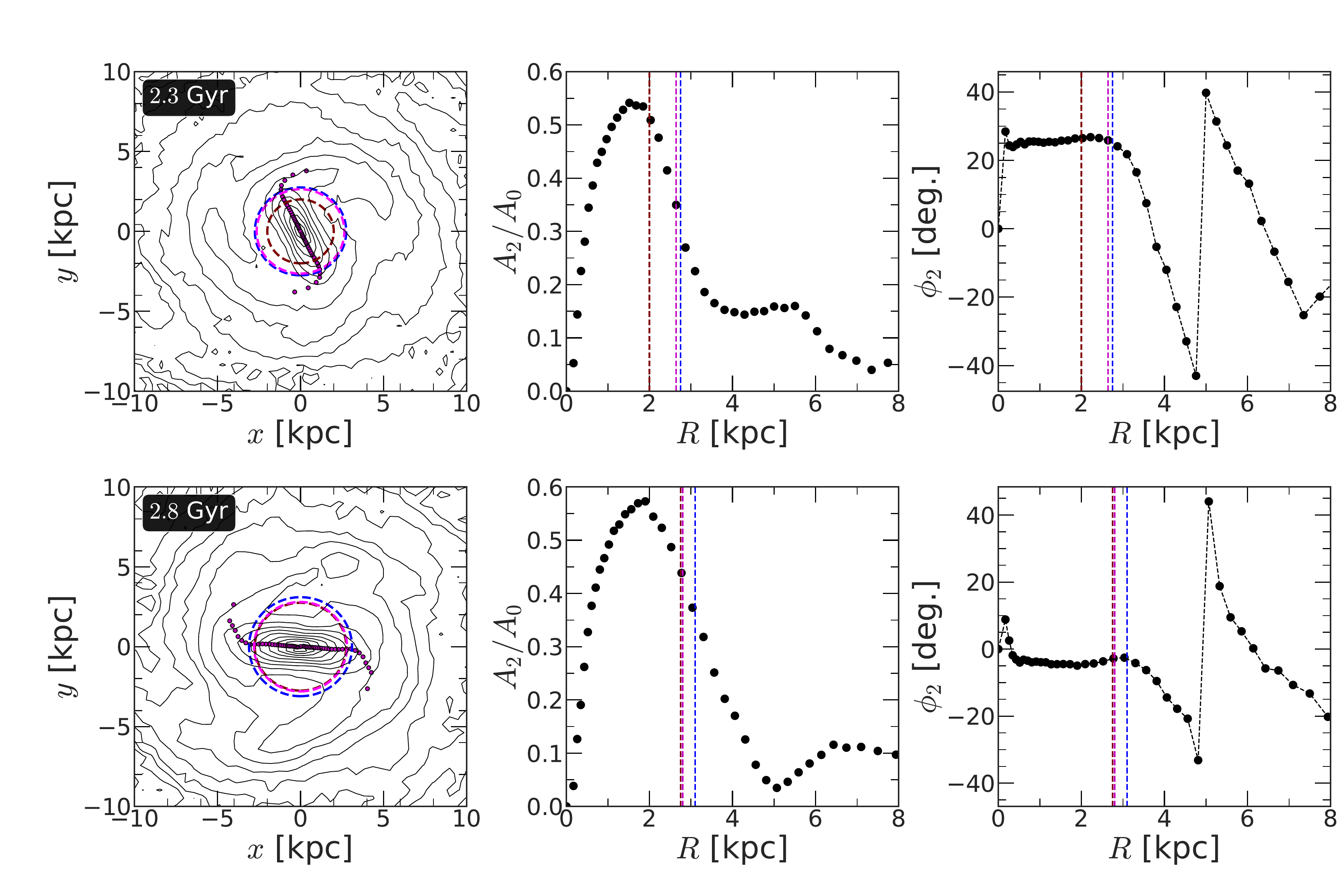}}
\caption{\textit{Effect of spirals in biasing the bar length.} \textit{Left panels:} Contours of constant surface density at two different times. Magenta points denote the orientation of the phase-angle of the $m=2$ Fourier moment. Dashed circles denote different bar length measurements.  \textit{Middle panels :} Radial variation of the $m=2$ Fourier coefficient (normalised by the $m=0$ component). Maroon dashed line denotes the value of $R_{\rm bar,2}$ whereas blue dashed line denotes the value of $R_{\rm bar,3}$. \textit{Right panels:} Radial variation of the phase-angle ($\phi_2$) of the $m=2$ Fourier moment. The magenta dashed line denotes the value of $R_{\rm bar,1}$. For further details, see the text. Presence of a spiral arm leads to an overestimation of bar length by $\sim 10-15$ percent in our fiducial bar model.}
\label{fig:bar_spiral_conundrum}
\end{figure*}

In a real galaxy, an $m=2$ bar is often accompanied by a spiral structure \citep[e.g. see][]{RixandZaritsky1995,Kruketal2018}. Past studies also postulated a bar-driven spiral formation scenario \citep[e.g.][]{Salo2010}. In a recent work, \citet{Hilmietal2020} studied in detail, the effect of spirals on inducing systematic overestimation in bar length measurement. The induced overestimation may vary in the range $15-55$ percent depending on the Hubble type of the galaxy and the strength of the spiral arms. However, the bar length estimators used in \citet{Hilmietal2020} depend on either the amplitude of the $m=2$ Fourier coefficient, or location of the dark-lanes. We revisit this issue here.
\par
First, we chose two times, namely, $t = 2.3 \Gyr$ and $2.8 \Gyr$ where our fiducial model harbours both a central bar and an outer spiral arm (see left panels of Fig.~\ref{fig:bar_spiral_conundrum}). It is well known that both an $m=2$ bar and a spiral would yield a non-zero value for the $m=2$ Fourier coefficient. In Fig.~\ref{fig:bar_spiral_conundrum} (see middle panels), we show the resulting radial profiles of the $m=2$ Fourier moment. As we can clearly see, the values of $A_2/A_0$ are non-zero in the outer parts due to presence of a spiral feature. However, we note that the corresponding phase angle ($\phi_{2}$) bears crucial information for deciphering an $m=2$ bar mode and an $m=2$ spiral. To illustrate, the phase angle of a bar remains (almost) constant over a range of radii (i.e. $d \phi_2/dR \sim 0$), which is one of its defining characteristics of an $m=2$ bar, whereas the $\phi_{2}$ varies systematically in the radial extent encompassing an $m=2$ spiral. Figure~\ref{fig:bar_spiral_conundrum} (see right panels) shows the such characteristic variations of the phase-angle ($\phi_2$) in a bar+spiral dynamical scenario. The radial distribution of $\phi_2$ shows a characteristic turn-over, matching the radial location where the spirals start to appear. To quantify any systematic overestimation induced by the presence of a spiral, we considered three bar length estimators, namely, $R_{\rm bar,1}$, $R_{\rm bar, 2}$, and $R_{\rm bar, 3}$ (for definition, see Sect.~\ref{sec:bar_measurement}). The resulting values of different bar length estimates are shown in Fig.~\ref{fig:bar_spiral_conundrum}. We found that when information about only the bar amplitude is taken into account, namely, $R_{\rm bar, 2}$, and $R_{\rm bar, 3}$, the resulting bar estimates are $\sim 5-15$ percent higher when compared with $R_{\rm bar,1}$ (which takes into account the information of $\phi_2$ variation). We note that the bar in our fiducial model is quite stronger as compared to the spirals. The overestimation of bar length (due to the presence of a spiral) would further increase in a dynamical scenario of `weaker bar and stronger spiral' so that the $m=2$ Fourier amplitudes for these two non-axisymmetric features are similar. Thus, the findings shown here, demonstrate that taking into account the information of phase-angle in measuring bar length can potentially alleviate the overestimation of bar lengths in presence of spirals. {\citet{Hilmietal2020} claimed that the bar length can fluctuate by up to 100 per cent on a dynamical timescale ($\sim 50-200 \Myr$). In our simulation, we did not did find such a strong fluctuation (on a dynamical timescale) in bar length measurement, especially towards the end of the simulation. A more detailed analysis towards understanding the origin and strength of this fluctuation is necessary, however, this is beyond the scope of this paper.}

\section{Comparison with bar length measurements from isophotal analysis}
\label{sec:comparison_photometry}

\begin{figure}
\centering
\resizebox{\linewidth}{!}{\includegraphics{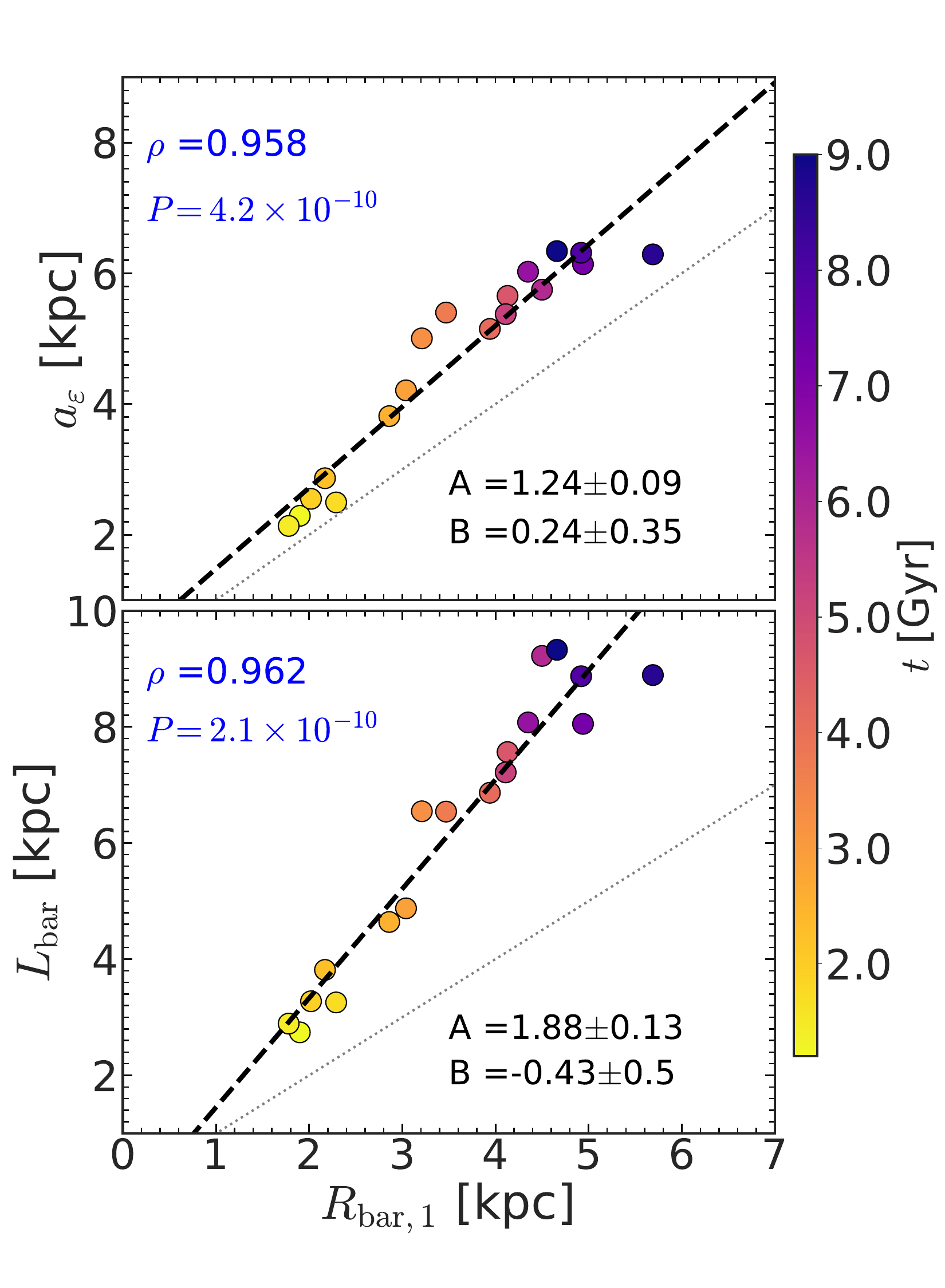}}
\caption{Correlation between the radial location of the peak ellipticity, $a_{\varepsilon}$ and the bar length, $R_{\rm bar,1}$ (obtained from the constancy of the phase-angle $\phi_2$) at different times (\textit{top panel}). Correlation between the photometric length of the bar, obtained from the constancy of position angle of fitted ellipses, $L_{\rm bar}$ and the bar length, $R_{\rm bar,1}$ (\textit{bottom panel}). In each case, a straight line of the form $Y = AX+B$ is fitted (black dashed line), and the corresponding best-fit values are quoted. The Pearson correlation coefficients are quoted in the top left corner of each sub-panel. The grey dotted line in each sub-panel denotes the one-to-one correspondence. Both the photometric estimators overestimate the length of the bar. }
\label{fig:photo_measurement}
\end{figure}

So far, we have considered different bar length estimators that make use of the Fourier amplitude (and phase-angle) obtained from the Fourier decomposition of intrinsic particle distribution. However, several past studies used different bar length estimators obtained from the isophotal analysis of de-projected surface brightness distribution \citep[e.g. see][]{Wozniaketal1995,Erwin2005,Aguerrietal2009}. Here, we carry out a comparative study between the bar length estimators from the isophotal analyses and the bar length estimators from the intrinsic Fourier analyses.
\par
For this work, we considered two bar length estimators (using information from the isophotal analysis), namely, $a_{\varepsilon}$ which denotes the location of the maximum ellipticity of the best-fit elliptical isophotes, and $L_{\rm bar}$, which denotes the radial extent within the position angle of the best-fit elliptical isophotes remain constant within $3-5 \degrees$ \citep[for details, see][]{Erwin2005}. To compute these quantities, we first performed {\sc {IRAF ellipse}} task on the face-on surface brightness maps. The resulting radial variation of the ellipticity ($\varepsilon = 1-b/a$, $a$ and $b$ being semi-major and semi-minor axes, respectively) and the position angle (PA) at $t = 1.95 \Gyr$ are shown in Fig.~\ref{fig:ellipsefitting_example}. We then measured the values of $a_{\varepsilon}$ and $L_{\rm bar}$ for $\sim 20$ snapshots covering the entire bar evolution scenario. The resulting variations of these quantities as well as their correlations with $R_{\rm bar,1}$ are shown in Fig.~\ref{fig:photo_measurement}. As seen from Fig.~\ref{fig:photo_measurement}, there are strong (positive) correlations between the $a_{\varepsilon}$  and $R_{\rm bar,1}$ as well as $L_{\rm bar}$  and $R_{\rm bar,1}$, as revealed by the values of the Pearson correlation coefficients. However, both the $a_{\varepsilon}$ and $L_{\rm bar}$ systematically overestimate the length of the bar when compared with $R_{\rm bar,1}$, as revealed from the slope values of the best-fit straight line (see Fig.~\ref{fig:photo_measurement}). This trend is in agreement with past studies \citep[e.g. see][]{Michel-Dansac2006,Petersenetal2023}

\section{Discussion}
\label{sec:discussion}

\subsection{Implication on determining the slow or fast bar}
\label{sec:implication_slowfast}

\begin{figure}
\centering
\resizebox{\linewidth}{!}{\includegraphics{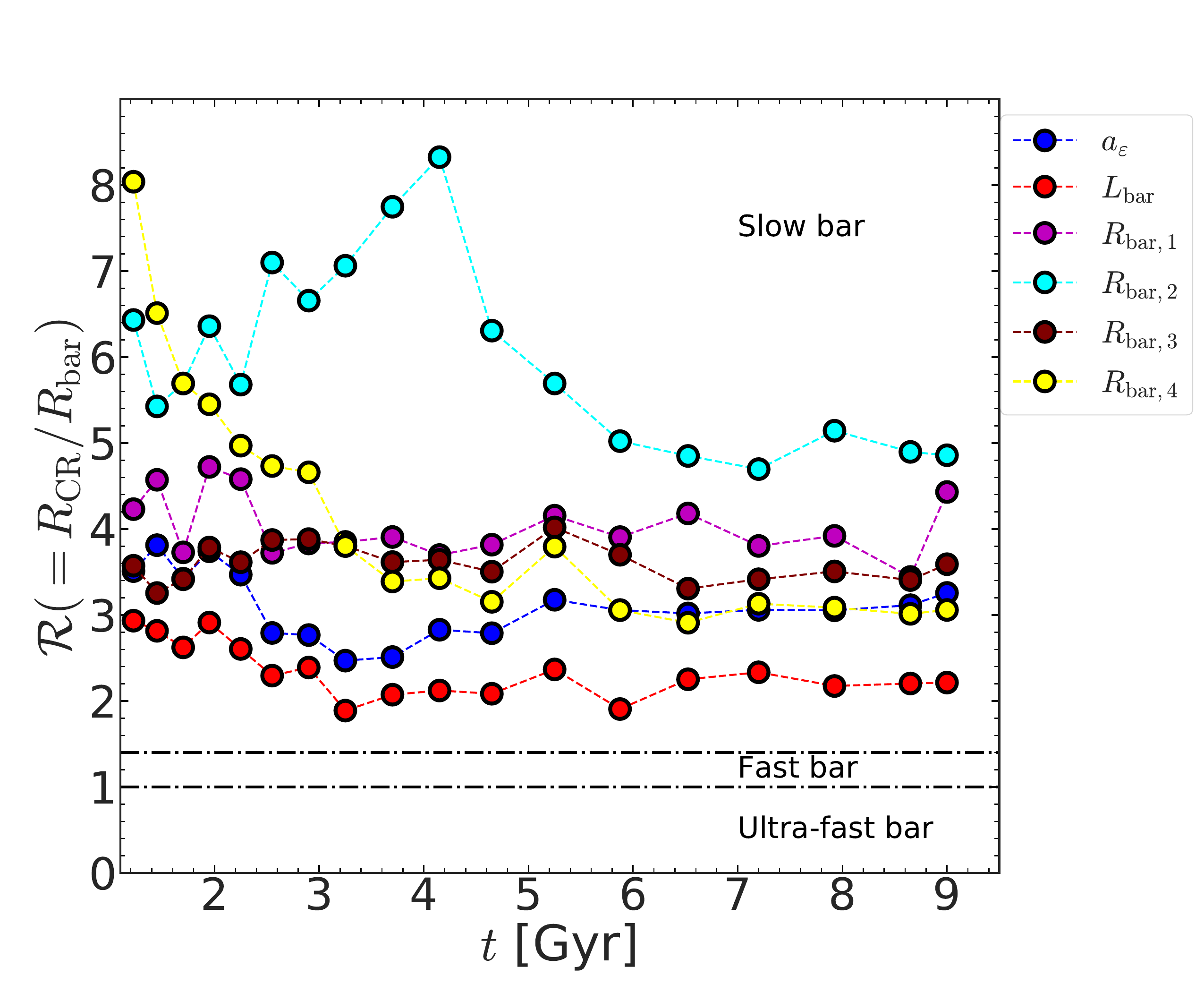}}
\caption{ Parameter $\mathcal{R}$, ratio of the corotation radius and the bar length,  as a function of time for the fiducial bar model. Different bar length measurements are used to compute the parameter $\mathcal{R}$ (see the legend). The horizontal dash-dotted lines denote the ultra-fast, fast, and slow bar regime.  Further details are provided in the text.}
\label{fig:implication_slowfast}
\end{figure}

One of the important applications of robust bar length measurement lies in the calculation of the parameter $\mathcal{R}$ which, in turn, determines whether a bar is classified as a fast or a slow bar (for further discussion, see Sect.~\ref{sec:Intro}). In other words, even if the pattern speed measurement and the corresponding location of corotation ($R_{\rm CR}$) are accurately measured, an erroneous estimate of bar length would lead to a wrong classification in the ultrafast-fast-slow bar regime \citep[e.g. see discussion in][]{Hilmietal2020,Cuomoetal2021}. 
\par
Here, we briefly test the robustness of the calculation of $\mathcal{R}$ parameter while using different bar length estimators (using information from the isophotal analysis and the Fourier decomposition of surface density distribution) as a function of time for our fiducial model. First, we calculated the rotation curve from the intrinsic particle kinematics by using the asymmetric drift correction \citep[see][]{BiineyTremaine2008}. One such example of rotation curve calculation at $t = 2.25 \Gyr$ is shown in Appendix~\ref{appen:vcirc_calc}. Then, using both the $\Omega_{\rm bar}$ measurement and the circular disc frequency ($\Omega_{\rm disc} = v_{\rm circ}/R$), we calculated the values of $R_{\rm CR}$ at different times. As for the bar length estimators, we choose six of them, namely, $R_{\rm bar,1}$, $R_{\rm bar,2}$, $R_{\rm bar,3}$, $R_{\rm bar,4}$, $a_{\varepsilon}$, and $L_{\rm bar}$. The resulting temporal variation of the parameter $\mathcal{R}$, obtained using different bar length estimates is shown in Fig.~\ref{fig:implication_slowfast} for our fiducial model. {As seen clearly from Fig.~\ref{fig:implication_slowfast}, the temporal evolution of $\mathcal{R}$ varies substantially depending on which bar length estimator has been used and the values of $\mathcal{R}$ can be overestimated by a factor of $\sim 1.5-2$, depending on the specific choice of a bar length estimator. This clearly outlines the importance of a robust, accurate bar length estimation while calculating the value of $\mathcal{R}$. We mention that the bar in our fiducial model always remains as a slow bar ($\mathcal{R} > 1.4$) and choice of any particular bar length estimator does not alter the fast and slow classification. However, in other simulated models (and for the observed galaxies) where a bar might shift from a fast rotator to a slow rotator (depending on the exact angular momentum transfer facilitating the slowing down of a bar) over time, an over-estimation of the bar length by a factor of $\sim 1.5-2$ (as shown here)  can potentially lead to identifying an otherwise fast and slow bar as an ultra-fast bar, in agreement with past studies \citep[e.g. see][]{Michel-Dansac2006,Hilmietal2020,Cuomoetal2021}.}


\subsection{Other issues}
\label{sec:other_issues}
In what follows, we discuss some of the implications and limitations of this work. First, the stellar bar in our fiducial model evolves self-consistently under the influence of a live dark matter halo -- and not in a rigid dark matter halo, as used in the models of \citet{Michel-Dansac2006}. Therefore, our model predicts the correct temporal evolutionary scenario for the $\mathcal{R}$ parameter, as compared to \citet{Michel-Dansac2006}. However, our fiducial model does not include any dissipative component, namely, the interstellar gas. Presence of such a dissipative component can have substantial dynamical effect on the overall generation and evolution of bar \citep[e.g. see][]{Bournaudetal2005}. It will be worth checking the correlations and robustness of different bar length estimators used here in numerical models of barred galaxies, which include the interstellar gas. 
\par
Secondly, we did not include the effect of dust extinction and the choices of different wavelength bands  in this work, while measuring the bar lengths via different estimators \citep[for further discussion, see e.g.][]{Michel-Dansac2006}. The central idea of this work has been to quantify any systematic intrinsic biases introduced by different bar length estimators and to study if there exists any well-defined correlation between these bar length estimators as the bar passes through its different evolutionary phases.
\par
 Furthermore, the fiducial model considered here grows quite a strong bar (which remains true for most of the time). It will be worth investigating the robustness of the scaling relations and the correlations reported here for a model which only harbours a weak bar. In addition, the total stellar mass in this model remains fixed and only the bar/T ratio increases as the bar grows in size and mass. It will be worth checking the robustness of the scaling relations and the correlations for a large sample of barred galaxies with different stellar mass. Similarly, the dark matter halo properties (e.g. central concentration, characteristic length, halo spin) have a vital impact on the bar formation and its subsequent evolution \citep[e.g.][]{Athanassoula2003,Dubinskietal2009,SahaandNaab2013}. The investigation of the robustness of the scaling relations and their universality under various dark matter halo configurations is beyond the scope of this paper; nevertheless it is of worth investigation.
Lastly, we computed the b/p length as a function of time for the fiducial model. However, we did not find any well-defined correlation between the bar length and the extent of the b/p structure.

\section{Summary}
\label{sec:conclusion}

In this work, we investigated the robustness of different bar length estimators while using an $N$-body model of a barred galaxy. In the model, the bar forms quite early ($\sim 1.1$ \Gyr), goes through a rapid growth phase, then it undergoes a vertical buckling instability, and, finally, it goes on to remain a strong bar at later times. In this work, we considered several bar length estimators, as commonly used in past literatures, and investigated if and how they lead to any systematic over(under)estimation of bar length in galaxies. The main findings are listed below.\\ 

\begin{itemize}
\item {$R_{\rm bar,2}$, obtained from identifying the location of the peak of the $m=2$ Fourier amplitude, systematically underestimates the bar length and this remains true for the entire bar evolutionary phase.}

\item {$R_{\rm bar,3}$, obtained from identifying the location where $A_2/A_0$ (the $m=2$ Fourier amplitude) drops to 70 percent of its peak value, provides reasonably accurate estimate for the the bar length. However, when a spiral arm is present, this method introduces systematic overestimation ($\sim 5-15$ percent) in the bar length measurement.}

\item{$R_{\rm bar,4}$, obtained from identifying the location where the ratio of bar and inter-bar intensity exceeds 2, systematically overestimates the bar length. This remains true for the entire bar evolutionary phase.}

\item{$R_{\rm bar,5}$, obtained using the criterion proposed in \citet{Aguerrietal2000}, {also provides reasonably accurate estimate for the the bar length. However, at later times when the radial profiles of the $m=2$ Fourier coefficient display multiple peaks within the extent of the bar, $R_{\rm bar,5}$ exhibits larger scatter with respect to the $R_{\rm bar,1}$ values.} }

\item{The strength dark-gaps (produced by bars), $\Delta \mu_{\rm max}$ correlates strongly with the bar length in early rapid growth phase and is only weakly anti-correlated during subsequent quiescent phase of bar evolution. However, the location of dark-gaps correlates only weakly with the bar length, hence, it cannot be used as a robust proxy for determining the bar length.}

\item{The presence of a spiral arm affects the bar length measurements, especially for those bar length estimators that rely solely on the amplitude of different Fourier moments and do not take into account the radial variation of the phase-angle of the $m=2$ Fourier moment.}

\item{Both the bar length estimators, $a_{\varepsilon}$, and $L_{\rm bar}$, obtained from the isophotal analysis, systematically overestimates the bar length and this remains true for the entire bar evolutionary phase.}
\end{itemize}

{The results shown here further demonstrate that depending on the specific choice of bar length estimators, the parameter $\mathcal{R}$ gets overestimated by a factor of $1.5-2$. The bar remains as a slow rotator (irrespective of the choice of bar length estimators) in our model throughout the simulation run, However, an overestimation of $\mathcal{R}$ by a factor of $1.5-2$ might lead to a potential misclassification in the slow or fast bar regime (along with other sources of uncertainty). This is in agreement with past findings.}

\section*{Acknowledgements}
{We thank the anonymous referee for useful comments which helped to improve this paper.} S.G. acknowledges funding from the Alexander von Humboldt Foundation, through Dr. Gregory M. Green's Sofja Kovalevskaja Award. This work has made use of the computational resources obtained through the DARI grant A0120410154 (P.I. : P. Di Matteo). S.G. thanks Vighnesh Nagpal for kindly rechecking the bar pattern speed calculation.  S.G. acknowledges the stimulus discussion during the `Galactic Bar, 2023' conference in Granada, Spain which served as a motivation for this work.

\bibliographystyle{aa.bst} 
\bibliography{my_ref.bib} 

\begin{appendix}

\section{Bar pattern speed and calculation of the circular velocity}
\label{appen:vcirc_calc}

In order to determine the parameter $\mathcal{R} (=R_{\rm CR}/R_{\rm bar})$, we need to measure the bar pattern speed ($\Omega_{\rm bar}$) and the rotation curve. These two quantities together provide the location of the corotation radius at a time, $t$. We measure the bar pattern speed ($\Omega_{\rm bar}$) by fitting a straight line to the temporal variation of the phase-angle ($\phi_2$) of the $m=2$ Fourier moment \citep[for details, see e.g.][]{Ghoshetal2022}. This assumes that the bar rotates rigidly with a single pattern speed in that time-interval. The resulting temporal evolution of the bar pattern speed ($\Omega_{\rm bar}$) is shown in Fig.~\ref{fig:bar_patternspeed}. As seen from Fig.~\ref{fig:bar_patternspeed}, the bar in our model slows down significantly during the entire course of evolution. 

Next, we briefly mention how the circular velocity is derived from the intrinsic particle kinematics. At a certain time, $t$, the circular velocity, $v_{\rm circ}$, is computed using the asymmetric drift correction \citep{BiineyTremaine2008}
\begin{equation}
v_{\rm circ}^2 (R) = v_{\phi}^2 (R)+ \sigma_{\phi}^2 (R) -\sigma_{R}^2 (R) \left(1 + \frac{\rm d \ ln \ \rho (R)}{\rm d \ ln \  R} + \frac{\rm d \ ln \ \sigma^2_R (R)}{\rm d \ ln \  R}\right)\,.
\label{eq:asy_drift1}
\end{equation}
\noindent Here, $\sigma_{R}$ and $\sigma_{\phi}$ denote the radial and the azimuthal velocity dispersion, respectively, and $v_{\phi}$ is the azimuthal velocity. One such example of the rotation curve, computed at $t = 2.25 \Gyr$ is shown in Fig.~\ref{fig:vcirc_t90}. Using the circular velocity as a function of radius, $R$ (equivalently, the rotation curve), we then compute the radial location of the corotation (CR), where the bar pattern speed ($\Omega_{\rm bar}$) commensurates with the underlying circular frequency of the disc stars, that is $\Omega (R = R_{\rm CR}) = \Omega_{\rm bar}$.

\begin{figure}
\centering
\resizebox{1.0\linewidth}{!}{\includegraphics{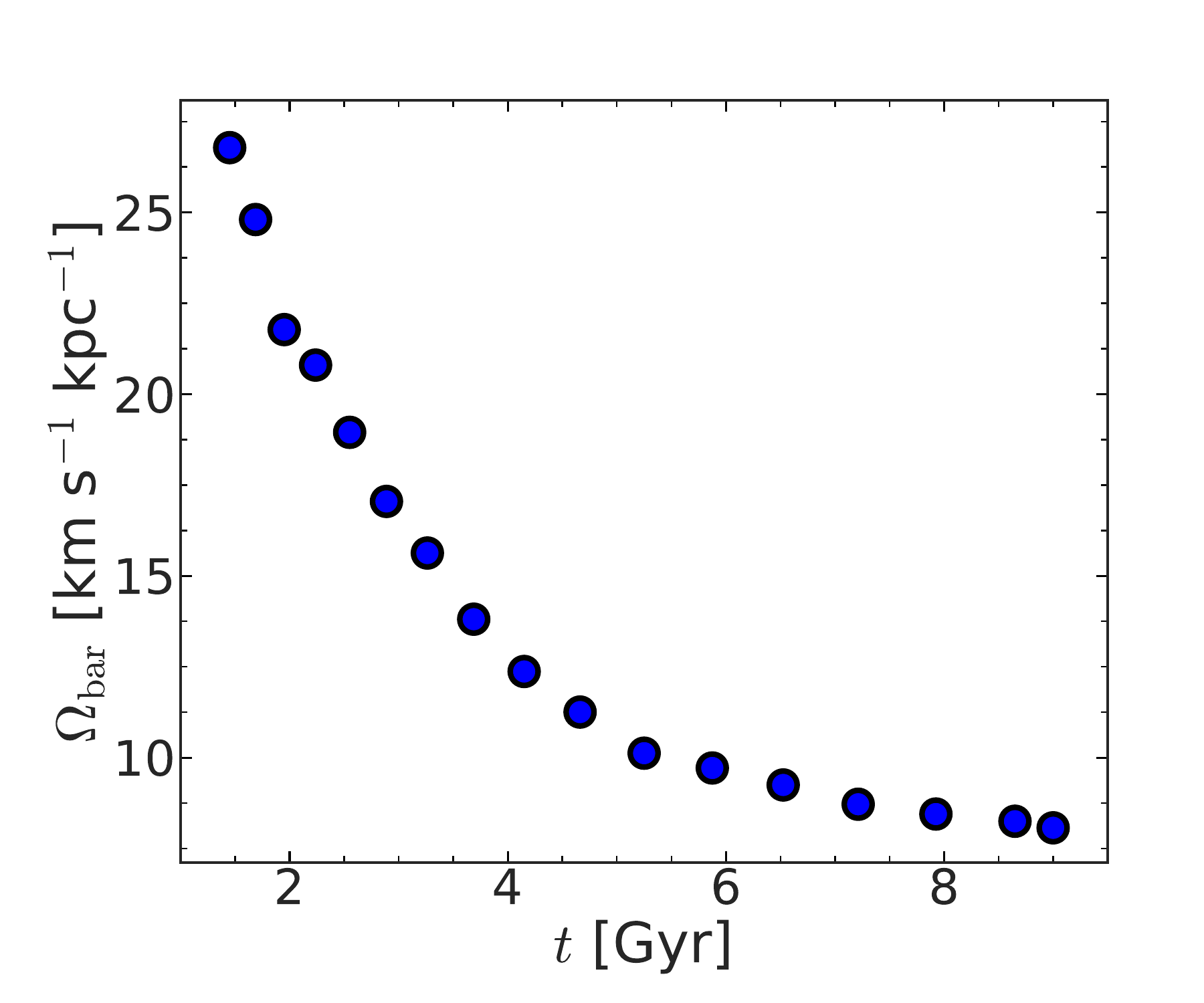}}
\caption{Temporal evolution of the bar pattern speed, $\Omega_{\rm bar}$ for the fiducial model. The bar pattern speed is calculated by measuring the temporal variation of the phase-angle ($\phi_2$) of the $m=2$ Fourier moment. For details, see the text in Appendix.~\ref{appen:vcirc_calc}.  The bar slows down substantially by the end of the simulation run ($t = 9 \Gyr$).}
\label{fig:bar_patternspeed}
\end{figure}

\begin{figure}
\centering
\resizebox{1.0\linewidth}{!}{\includegraphics{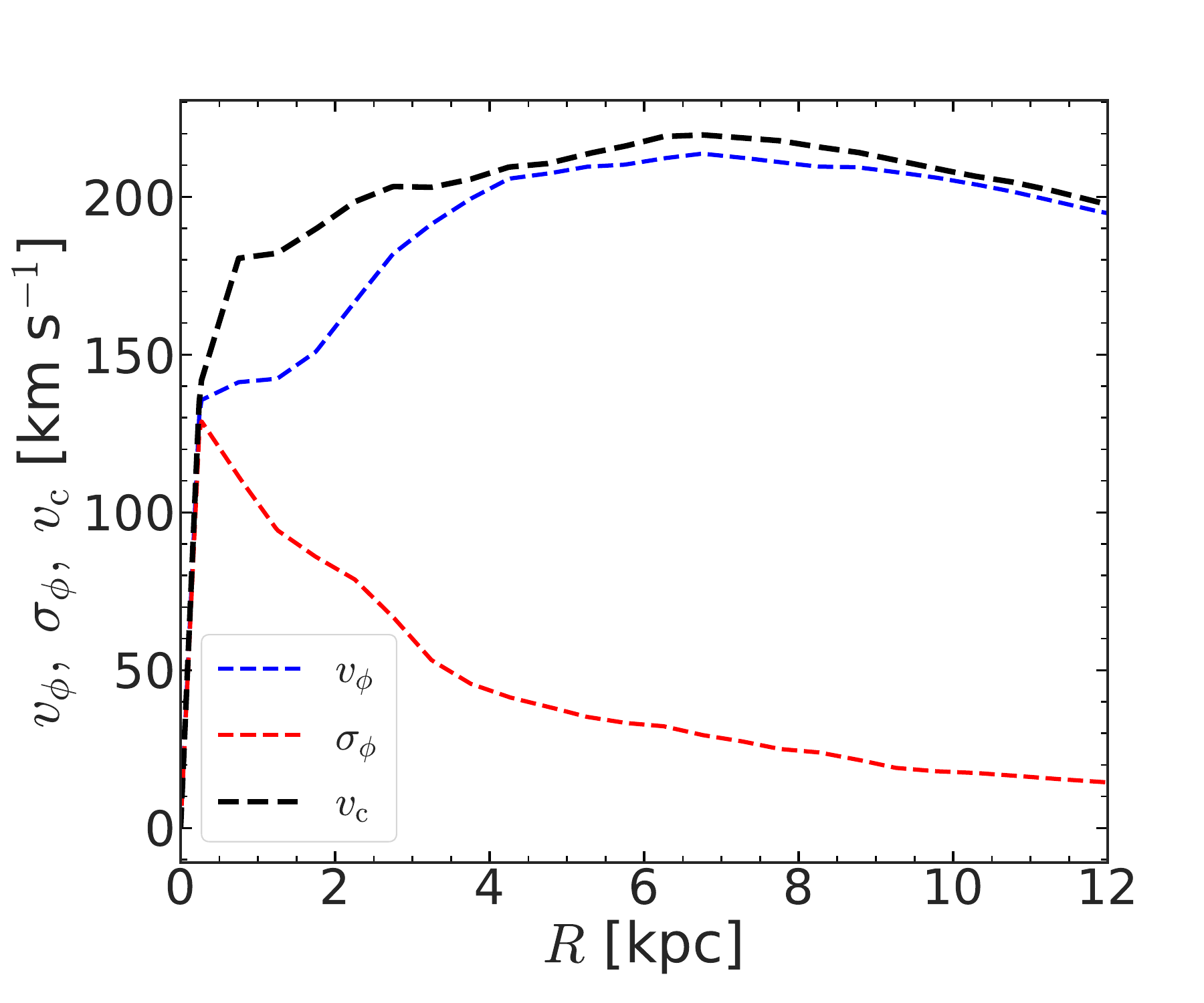}}
\caption{Radial variation of the circular velocity, $v_{\rm circ}$, calculated using the asymmetric drift correction (Eq.~\ref{eq:asy_drift1}) at $t = 2.25 \Gyr$.}
\label{fig:vcirc_t90}
\end{figure}

\section{Multiple peaks in the $m=2$ Fourier coefficient \& robustness of $R_{\rm bar,5}$ }
\label{appen:dissection_rbar5}

\begin{figure}
\centering
\resizebox{1.0\linewidth}{!}{\includegraphics{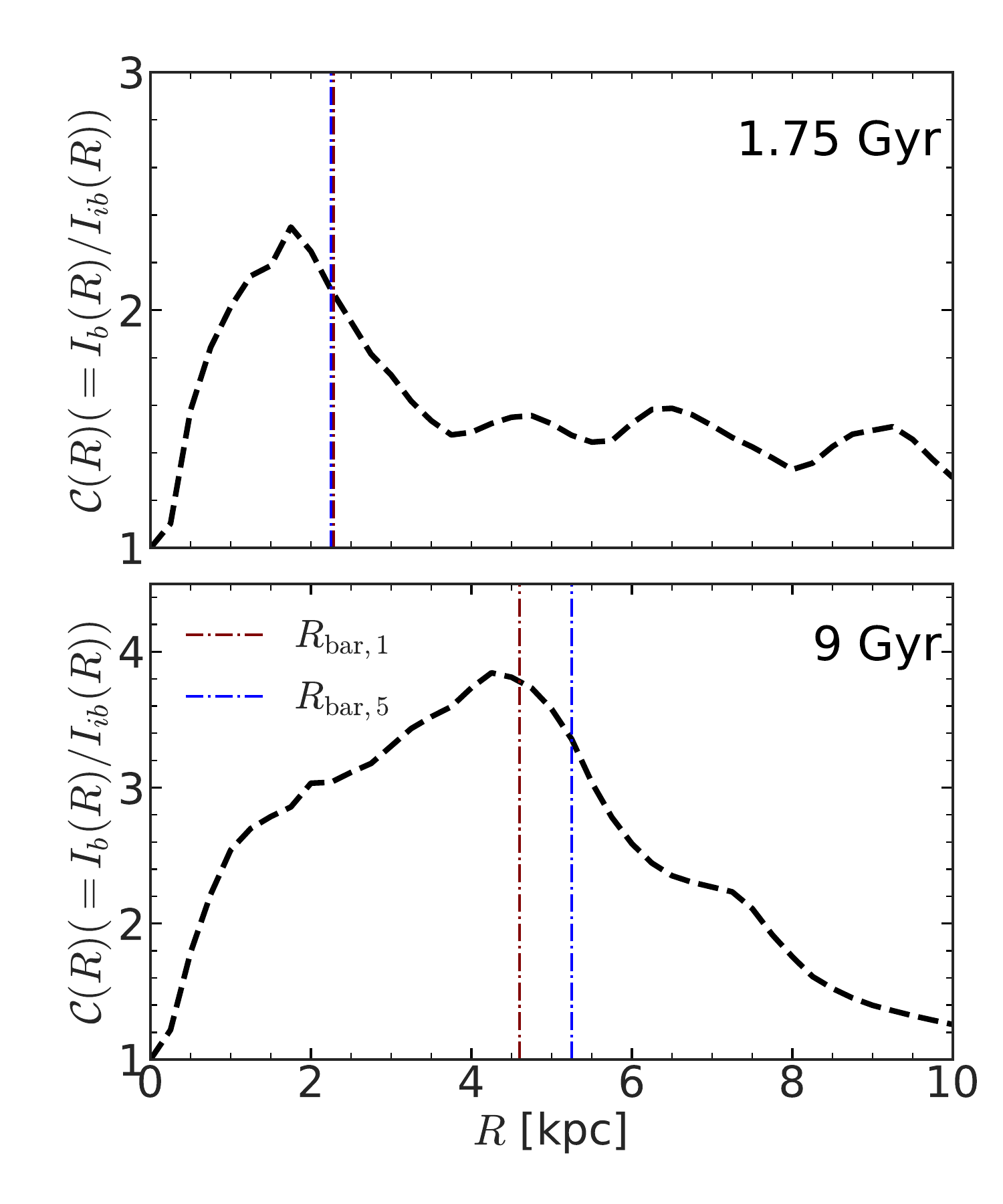}}
\caption{{Radial variation of the contrast profile, $\mathcal{C}(R) (= I_b (R)/I_{ib}(R))$ at  $t = 1.75 \Gyr$ (corresponding to an early growth phase of the bar) and $t = 9 \Gyr$ (corresponding the to end phase of the bar). The vertical maroon dashed line denotes the value of $R_{\rm bar,1}$, while the vertical blue dashed line denotes the value of $R_{\rm bar,5}$. At $t = 9 \Gyr$, the contrast profile, $\mathcal{C}(R)$ shows multiple within the extent of the bar as opposed a single-peaked contrast profile at $t = 1.75 \Gyr\ $:}}
\label{fig:dissection_rbar5}
\end{figure}

{Figure~\ref{fig:bar_length_measure} shows a larger degree of scatter for the $R_{\rm bar,5}$ measurement at later times ($t > 5 \Gyr$). Here, we investigate this in further detail. To do so, we first chose two time-steps, namely, $t = 1.75 \Gyr$ (corresponding to an early growth phase of the bar) and $t = 9 \Gyr$ (corresponding the to end phase of the bar). Then, we computed the contrast profiles $\mathcal{C}(R) (= I_b (R)/I_{ib}  (R))$ for these two selected time-steps. This is shown in Fig.~\ref{fig:dissection_rbar5}. As seen clearly, at $t = 9 \Gyr$, the contrast profile, $\mathcal{C}(R)$ shows multiple peaks within the extent of the bar as opposed a single-peaked contrast profile at $t = 1.75 \Gyr$.
We mention that the bar in our fiducial model is quite strong and, more importantly, it has gone through a strong vertical buckling instability to create a face-on b/p structure at subsequent times \citep{Ghoshetal2023b}. As a result of this, at later times, the radial profiles of the $m=2$ Fourier coefficients display three radially well separated peaks (see Fig.~\ref{fig:fourier_radial_profiles}), similar to what was shown in \citet{Vynatheyaetal2021}. This, in turn, makes the contrast profile $\mathcal{C}(R)$ multi-peaked within the extent of the bar at later times. Furthermore, the underlying assumption behind the $R_{\rm bar,5}$ measurement is that the contrast profile $\mathcal{C}(R)$, within the bar extent, can be well approximated by a Gaussian profile \citep[for details, see][]{Aguerrietal2000}. However, the contrast profiles $\mathcal{C}(R)$, computed at later times for our model (when the model harbours a face-on b/p structure), are not well approximated by a Gaussian profile; thereby, leading to an inconsistent bar length measurement by this technique. This highlights the limitation of this bar length estimator in cases where the bar is strong enough to form multiple sub-structure within the bar, characterised by the presence of multiple peaks in the radial $m=2$ Fourier coefficient \citep[for further details, see][]{Vynatheyaetal2021}.}

\section{Isophotal analysis of face-on surface brightness distribution}
\label{appen:surface_photometry}

In order to compute the values of $a_{\varepsilon}$, and $L_{\rm bar}$ (for definitions, see Sect.~\ref{sec:comparison_photometry}), first we constructed the face-on surface brightness map from the intrinsic particle distribution (see panel (a) in Fig.~\ref{fig:darklanes}) and then we performed the {\sc {IRAF ellipse}} task to obtain the radial profiles of the ellipticity ($\varepsilon = 1-b/a$, $a$ and $b$ being semi-major and semi-minor axes, respectively) and the position angle (PA). One such example in shown in Fig.~\ref{fig:ellipsefitting_example}. Within the extent of the bar, the radial profile of ellipticity shows a clear peak and the PA remains almost constant within the bar extent.

\begin{figure*}
\centering
\resizebox{1.0\linewidth}{!}{\includegraphics{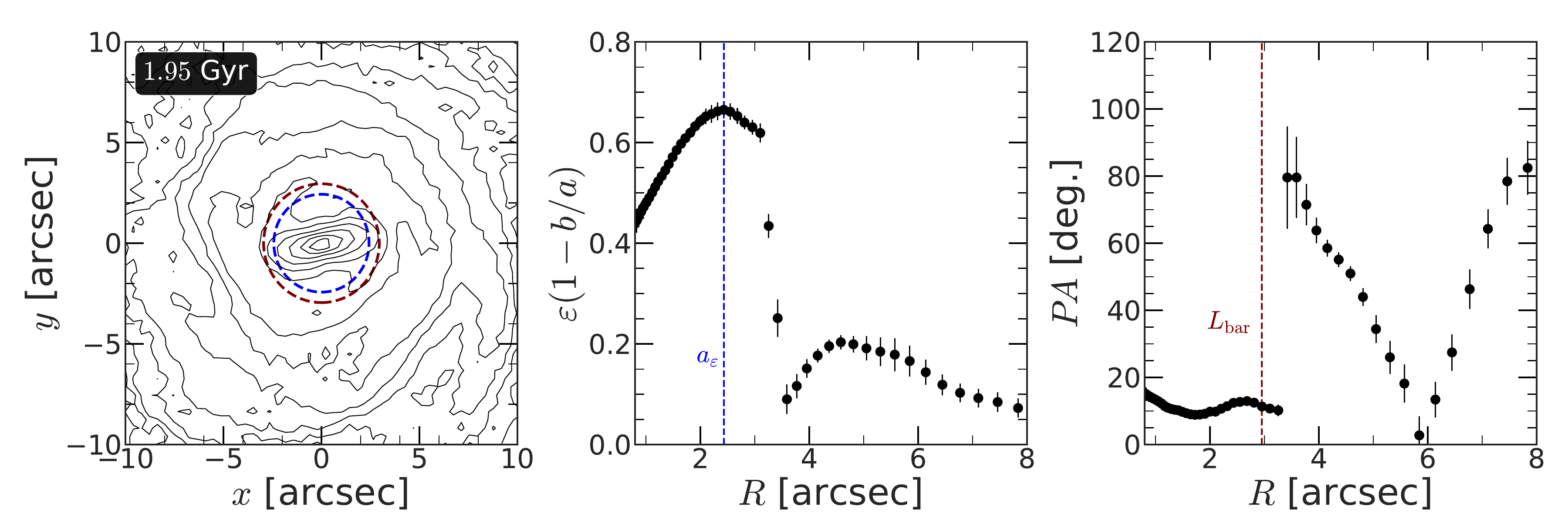}}
\caption{Measuring the bar length from photometric analysis. \textit{Left panel:} Contours of constant surface brightness at $t = 1.95 \Gyr$. The blue and the maroon circles denote the extents of $a_{\varepsilon}$ and $L_{\rm bar}$, respectively. \textit{Middle panel: } Radial profiles of the ellipticity ($\varepsilon = 1-b/a$), where $a$ and $b$ are semi-major and semi-minor axes. The vertical blue dashed line denotes the radial location of $a_{\varepsilon}$. \textit{Right panel:}  Radial profiles of the  position angle (PA). The vertical maroon dashed line denotes the radial location of$L_{\rm bar}$. Here, 1 arcsec = $1 \kpc$. The radial profile of ellipticity shows a clear peak, and the PA remains almost constant within the bar extent.}
\label{fig:ellipsefitting_example}
\end{figure*}

\end{appendix}

\end{document}